\documentclass[aps,prx,twocolumn,floats,showpacs,superscriptaddress,nofootinbib]{revtex4-1}
\usepackage{graphicx,epsfig}
\usepackage{times,bbm}
\usepackage{graphics,dcolumn,bm,float}
\usepackage{amssymb,amsmath,rotate,color}
\usepackage[title,titletoc,toc]{appendix}
\usepackage{mathtools}
\usepackage{booktabs}
\usepackage{tcolorbox}
\usepackage[pagebackref=false,colorlinks,linkcolor=magenta,citecolor=blue,urlcolor=magenta]{hyperref}

\usepackage{wrapfig}
\usepackage{lipsum}
\usepackage{mwe}
\usepackage[mathlines]{lineno}
\usepackage{hyperref}
\usepackage{breqn}
\usepackage{bbold}
\usepackage{pgfplots}
\usepackage{float}
\usepackage{tkz-euclide}
\usepackage{braket}
\usepackage{physics}
\usepackage[caption=false]{subfig}
\usepackage[export]{adjustbox}
\usepackage{tikz}
\usetikzlibrary{through,calc}
\usetikzlibrary{positioning}

\newcommand{\be}{\begin{equation}}
\newcommand{\ee}{\end{equation}}

\newcommand{\bea}{\begin{eqnarray}}
\newcommand{\eea}{\end{eqnarray}}
\newcommand{\nn}{\nonumber}

\newcommand{\eps}{\varepsilon}

\newcommand{\bs}{\boldsymbol}
\newcommand{\bmt}{\left[\begin{matrix}}
\newcommand{\emt}{\end{matrix}\right]}

\begin{document}
\preprint{}
\title{Perpendicular Andreev reflection: Solid state signature of black hole horizon}

\author{Z. Faraei}
\email{zahra.faraei@gmail.com}
\affiliation{Department of Physics$,$ Sharif University of  Technology$,$ Tehran 11155-9161$,$ Iran}
\affiliation{Abdus Salam ICTP$,$ Strada Costiera 11$,$ I-34151 Trieste$,$ Italy}

\author{S.A. Jafari}
\email{jafari@sharif.edu}
\affiliation{Department of Physics$,$ Sharif University of  Technology$,$ Tehran 11155-9161$,$ Iran}
\affiliation{Perimeter Institute For Theoretical Physics$,$ 31 Caroline St. N.$,$ Waterloo$,$ ON$,$ N2L 2Y5$,$ Canada}

\date{\today}

\begin{abstract}
Beenakker noticed that the peculiar band structure of Dirac fermions in 2D solids allows for the specular Andreev reflection in these systems
which has no analogue in other 2D electron systems. An interesting deformation of the Dirac equation in the solid state
is to tilt it which has now materials realization. 
In this work we report another peculiar feature of the Andreev
reflection in tilted 2D Dirac cone systems. The tilt deformation of the Dirac equation is characterized by two
parameters ${\bs\zeta}=\zeta(\cos\theta,\sin\theta)$. 
We show that when the tilt parameter is tuned to its "horizon value" $\zeta=1$, irrespective of the incidence angle of electrons, the
Andreev reflected hole is always reflected {\em perpendicular} to the interface. Furthermore at the horizon value, if the
tilt direction is perpendicular to the interface, the conductance will be energy independent for the entire 
sub gap energies. For generic tilt $\bs\zeta$, the tilt dependence of the conductance line shape can be used to extract information about 
the direction and magnitude of the tilt. 
\end{abstract}



\maketitle
\narrowtext

\section{Introduction}
Dirac and Weyl equations by now have become  ubiquitous in solids~\cite{Fuseya,Armitag,CastroNetoGraphene}.
But since solids are mounted on lattices, the resulting deformation of the Dirac theory disobeys the Lorentz symmetry. 
One interesting deformation of the Dirac equation in solids is to tilt it~\cite{Beenakker2016KleinTunneling,Varykhalov2017,Lozovik,Soluyanov2015,Pyrialakos2017}. 
The tilted Dirac fermions were originally reported in organic material $\alpha$-(BEDT-TTF)$_2$I$_3$~\cite{Suzumura2006,Tajima2006}.
Recently it has been
proposed that on certain nonsymmorphic lattices, such a tilt can be manipulated by perpendicular electric fields~\cite{tohid}. 
Even though the Dirac cone in pristine graphene is not tilted,
the strain pattern can impose small tilt in the spectrum of graphene~\cite{Cabra2013}.

From a solid state physicist perspective, the tilt deformation mixes the energy and momentum, thereby deforming 
the circular or spherical Fermi surfaces into elliptical or ellipsoidal Fermi surfaces. 
From this point of view, the tilt deformation of the Dirac/Weyl equation 
leads to clear signatures in various spectroscopies in two and three dimensional materials.
Examples include the conductance~\cite{Beenakker2016KleinTunneling,Jalil2017}, spin transport~\cite{Sinha}, Klein tunneling~\cite{Charlier}, 
anomalous Hall conductivity~\cite{Zyuzin}, magnetotransport~\cite{sharma}, 
plasmon excitations~\cite{sahar_kink,Goerbig1} and optical response~\cite{Carbotte1,Carbotte2,experimental,lee},
as well as in the pairing energy scales~\cite{Alidoust2017,Alidoust2019evolution}.

From a more fundamental perspective, mixing the energy and momentum can be alternatively
viewed as mixing of time and space coordinates. Such a mixing changes the future "light cone" of
the electrons or holes living in such solids~\cite{Nissinen2017,Volovik_2018,tohid}. 
The tilt is parameterized by a tilt parameter $\zeta=v_t/v_F$ which is the ratio between two velocity scales: $v_t$ specifies
how much the Dirac cone is tilted, and $v_F$ determines the major velocity scale associated with the solid angle 
subtended by the Dirac cone in energy-momentum space. In the geometric language the condition $\zeta=1$ 
marks the black hole horizon with which a Hawking radiation is associated~\cite{VolovikBH,Meng2018BH,tohid,JafariEB}. 
In the standard solid-state language, the $\zeta=1$ marks a Lifshitz transition
across which the superconducting transition temperature is enhanced~\cite{Volovik_2018,Alidoust2017}. 
By increasing the tilt parameter from $\zeta=0$ to $\zeta=1$, the circular Fermi surfaces of the
two-dimensional Dirac system will evolve into ellipses which tend to a line-segment connecting the 
two Dirac nodes in $\zeta=1$ limit. To this extent the density of states at the Fermi level 
will be enhanced which in turn gives rise to the enhanced pairing correlations at $\zeta=1$. 
This applies to the uniform tilt parameter $\bs\zeta$. The geometric point of view becomes
particularly useful when one allows $\bs\zeta$ to depend on coordinates~\cite{tohid}. In such
situation, the $\zeta=1$ corresponds to an event horizon,
one expects an observer at horizon to disagree on the particle content of a state coming from
$\zeta=0$ part of the spacetime. By basic uncertainty principle, such an increase in $\Delta N$ will
reduce the uncertainty in the phase, $\Delta\phi$. Therefore to that extent, approaching $\zeta=1$ 
is expected to enhance superconducting correlations. So if the tilt can depend on coordinates,
in addition to density of states effects, there can be additional pairing correlations coming 
from the curvature of the spacetime felt by electrons~\cite{tohid,JafariEB,jalali2019polarization}. 
This motivates us to study Andreev processes -- which are hallmark of superconducting states -- in tilted Dirac fermions
and investigate their evolution as a function of the tilt parameter $\zeta$. 

\begin{figure}[t]
\centering
\includegraphics[width=7cm]{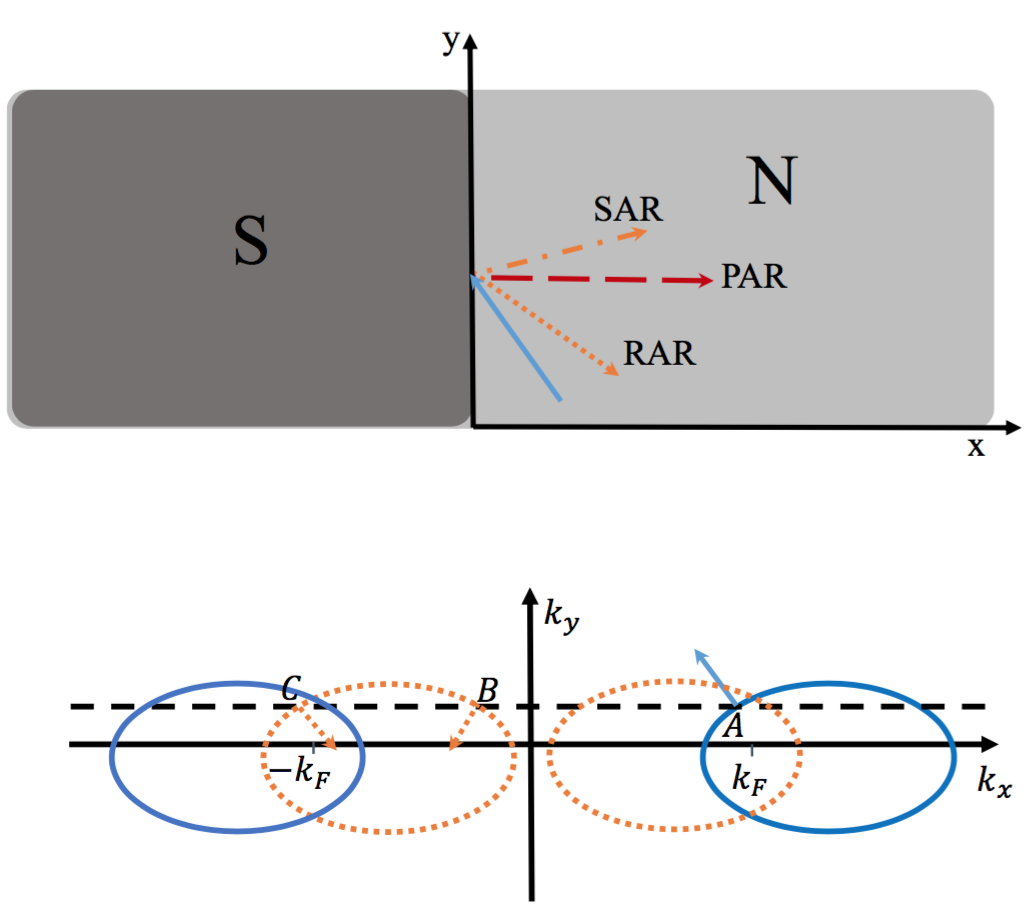}
\caption{(Color online) Top panel shows schematic representation of the S$|$N junction and three
types of Andreev reflection. Bottom panel shows the electron (solid) and hole (dotted line) Fermi surfaces
in two valleys. Arrows indicate the gradients and hence velocities. See text for explanation. 
The right and left Fermi surfaces correspond to right and left valleys. In the absence of tilts, 
the electron and hole Fermi surfaces coincide. 
}
\label{schematic.fig}
\end{figure}

Beenakker has found an interesting from of Andreev reflection~\cite{Andreev} (AR) which can only occur under specific
circumstances in Dirac materials~\cite{Beenakker_SAR}. As depicted in Fig.~\ref{schematic.fig}, when an electron in a normal material (blue arrow at A) 
hits a superconductor, it has two options: either to get reflected as an electron or as a hole (dotted orange arrow). 
When the Fermi energy is large the schematic drawing in the bottom panel of Fig.~\ref{schematic.fig} is relevant. 
The velocity is the gradient of the Fermi surface which in the case of electron (blue) Fermi surface is outward and for the
holes (orange dotted surface) is inward. The black dashed line is a constant $k_y$ line. The solution with $k_x$ at B point
does not correspond to (Andreev) reflected hole, while the one at C is Andreev reflected hole and traverses the path
opposite to the incident electron. This is the standard retro-AR (RAR) and takes place in the interface of any 
normal conductor having extended Fermi surface with a superconductor. 
What Beenakker noticed
was that in Dirac materials when the energy $\eps$ at which we are measuring is much larger than $E_F$, the 
holes can flip their helicity. This corresponds to the reversal of the direction of the dotted orange arrows
at B and C in Fig.~\ref{schematic.fig}. In this way, Beenakker found that the reflected hole can also be specular and dubbed it specular AR (SAR).
The sign conventions in Fig.~\ref{schematic.fig} are such that
in SAR (RAR) the sign of the angle of the reflected hole is the same as (opposite to) the sing of the angle of incidence of the 
incident electron. 
The magnitude of the angle of the reflected hole is controlled by the ratio $\eps/E_F$. 
From the above argument it is clear that the above ratio determines whether we are in RAR or SAR regime. 
The $\eps/E_F\gg 1$ is SAR dominated while in the $\eps/E_F\ll 1$ regime we are dealing with a big
Fermi surface and therefore the RAR is dominant AR process. 

In this paper, we show that when the tilt deformation is added to a 2+1 dimensional Dirac theory,
in addition to SAR and RAR, there is another type of AR in which the reflected hole angle 
is always $\pi/2$, independent of the incident electron angle. This takes place at $\zeta=1$
and means that the hole is always reflected back perpendicular to the interface. 
In this paper we call this type of AR, the perpendicular AR (PAR). The generic effect of non-zero $\zeta$
will be to bring the reflection angle of the reflected holes closer to perpendicular direction. The $\zeta=1$
will be an extreme case where the Andreev reflected hole always returns perpendicular to the interface. 
To see how this happens in Fig.~\ref{schematic.fig}, note that $\zeta$ is basically the eccentricity of the
elliptic Fermi surfaces. In the present figure, where the direction of tilt is assumed to be along $x$ direction,
in the $\zeta\to 1$ limit, the ellipse-shaped Fermi surface will reduce to a line segment. Therefore the only 
possible solutions will correspond to $k_y=0$ and the B and C points will lie on $k_y=0$ line which correspond to PAR. 

The paper is organized as follows: In section~\ref{formulation.sec} we formulate the AR 
for tilted Dirac cone systems as a natural extension of the AR in graphene~\cite{Beenakker_SAR}. 
In section~\ref{results.sec} we explore the dependence of AR in two regimes of RAR and SAR on
details of the magnitude and direction of the tilt and present peculiar conductance line-shapes
controlled by the tilt parameter $\bs\zeta$. We end the paper by a summary and discussion in section~\ref{discuss.sec}.

\section{Andreev reflection in tilted Dirac fermion systems}
\label{formulation.sec}
In order to build on the work of Beenakker~\cite{Beenakker_SAR} and to be able to 
compare our results against his results, let us deform the Dirac theory of graphene
by a tilt deformation. So we consider a generic tilt parametrized by  $\bs\zeta=\zeta_x \hat{x} + \zeta_y \hat{y}$ which 
tilts the Dirac cones along the angle $\tan^{-1}( \zeta_y/ \zeta_x)$ relative to $k_x$ axis.
This tilt deformation of the Dirac equation introduces a new term into the single particle Hamiltonian of Graphene
which is proportional to unit matrix $\sigma_0$ as follows,
\bea
\nn
H&=&\begin{pmatrix}
H_+ & 0 \\
0 & H_-
\end{pmatrix},\\
\nn
H_\pm &=& - i \hbar v (\sigma_x \partial_x \pm \sigma_y \partial_y) +U\\
& \mp&   i \hbar v \sigma_0(\zeta_x \partial_x + \zeta_y \partial_y ).
\label{tgraphene.eqn}
\eea
The $4\times 4$ matrix Hamiltonian operates on the space of  
four-component spinors $(\psi_{A+}, \psi_{B+}, \psi_{A-}, \psi_{B-})$. The indices $A$ and $B$ label
the two sublattices of the honeycomb lattice of carbon
atoms, while the indices $\pm$ label the two valleys of the
band structure. There is an additional spin degree of
freedom, which in the absence of spin-orbit interaction does not appear in the Hamiltonian.
The $2 \times 2$ Pauli matrices $\sigma_i$ act on the sublattice index.
Here $v$ is the Fermi velocity and $U$ is an externally applied electrostatic
potential. 

We consider a sheet of Graphene in the $xy$ plane.
As depicted in upper panel of Fig.~\ref{schematic.fig}, the half-space $x < 0$ is superconducting, 
while the region $x > 0$  is in the normal state. 
Eq.~\eqref{tgraphene.eqn} can be extended to Nambu space to include the
superconducting correlations. 
The electron and hole excitations are described by the Bogoliubov-De Gennes equation,
\be
\begin{pmatrix}
H-E_F  &  \Delta\\
\Delta^\dag  &  E_F - {\cal T} H {\cal T}^{-1}
\end{pmatrix}
\begin{pmatrix}
u\\
v
\end{pmatrix}
=
\eps
\begin{pmatrix}
u\\
v
\end{pmatrix},
\ee
where $u$ and $v$ are the electron and hole wave functions,
$\eps> 0$ is the excitation energy (relative to the 
Fermi energy $E_F$),  and  the time-reversal (TR) operator is
${\cal T}=\tau_x\sigma_z {\cal K}$, with ${\cal K}$ the complex conjugation operator. 
The $\tau_x$ is meant to exchange the valley indices. Since the $\sigma$ refers to 
sublattice index, the $\sigma$ part of the TR is represented by $\sigma_z$~\cite{Ando}. 
Eq.~\eqref{tgraphene.eqn} of the tilted Dirac cone in graphene is constructed in such a way
that it is TR invariant, namely, ${\cal T} H {\cal T}^{-1} = H$.
The pair potential $\Delta$ couples time-reversed electron 
and hole states, which can be considered as a step function at 
$x=0$. This assumption is valid if the superconducting coherence length in the superconducting region (S) 
is much smaller than the Fermi wave length in the normal region (N). 

The BdG equation will be straightforward generalization of the untilted graphene, 
and gives two decoupled sets of equations of the form,
\begin{widetext}
\be
\begin{pmatrix}
\tau (\zeta_x k_x + \zeta_y k_y ) - E_F & k_x-\tau i k_y & \Delta_0 e^{i\phi}  & 0\\
k_x +\tau i k_y & \tau(\zeta_x k_x + \zeta_y k_y ) - E_F & 0 & \Delta_0 e^{i\phi}\\
\Delta_0 e^{-i\phi}  & 0 & E_F -\tau (\zeta_x k_x + \zeta_y k_y ) & -(k_x -\tau i k_y)\\
0 & \Delta_0  e^{-i\phi} &  -(k_x +\tau i k_y) & E_F -\tau  (\zeta_x k_x + \zeta_y k_y ) 
\end{pmatrix}
\begin{pmatrix}
\psi_{A+}\\
\psi_{B+}\\
\psi_{A-}^*\\
-\psi_{B-}^*
\end{pmatrix}
=
\eps_{\tau}
\begin{pmatrix}
\psi_{A+}\\
\psi_{B+}\\
\psi_{A-}^*\\
-\psi_{B-}^*
\end{pmatrix}
\ee
\end{widetext}
where  $\tau=\pm$ labels the valley index. 
For any given value of $\tau$ there are four eigenvalues for the
above equation which are given by,
$\pm \sqrt{\Delta_0^2+( E_F -\tau  {\bs\zeta}.{\bs k} \pm k )^2}$ 
with $k^2=k_x^2+k_y^2$. 
The eigenvectors are the same as untilted case~\cite{Beenakker_SAR} as the tilt perturbation in Eq.~\eqref{tgraphene.eqn}
is proportional to unit matrix $\sigma_0$ and does not alter the eigenvectors.
The only difference in the eigenfunctions of the tilted case with respect to the upright case is that
one has to perform the replacement $E_F\rightarrow E_F  -\tau {\bs\zeta}.{\bs k}$.
This replacement carries over to all quantities derived from the eigenfunctions. 
Therefore, at a given energy $\eps$ and corresponding to wave vector ${\bs k}=(k_x,k_y)$ 
the angle of incidence of the electron is given by,
\bea
\alpha=\arcsin \big\{ \hbar v k_y / [\eps+E_F - \tau  {\bs\zeta}.{\bs k}] \big\}.
\label{alfa.eqn}
\eea  
Similarly the reflection angle of the hole will be,
\bea
\alpha'=\arcsin \big\{ \hbar v k_y / [\eps- E_F + \tau  {\bs \zeta}.{\bs k'}] \big\},
\label{alfa'.eqn}
\eea  
where ${\bs k'}=(k'_x,k_y)$. Note that $k_y$ is conserved due to transnational invariance along the border $x=0$ separating S and N regions.
The $x$ components of the wave vectors of the electron and the reflected hole are,
\bea
& k_x= \big\{ \hbar v k_y / [\eps+E_F - \tau  {\bs\zeta}.{\bs k}] \big\} \cos \alpha,\nn\\
& k'_x= \big\{ \hbar v k_y / [\eps- E_F + \tau  {\bs\zeta}.{\bs k'} ] \big\} \cos \alpha'.
\label{k.eqn}
\eea
 Substituting Eq.~\eqref{k.eqn} in Eq.~\eqref{alfa.eqn} and~\eqref{alfa'.eqn}, results in  an equation between 
$\sin \alpha$ and $\sin \alpha'$:
\bea
\nn
&&\{\zeta_x^2  f^2(\alpha) +[1+\tau\zeta_y f(\alpha) ]^2\} \cos^2 \alpha' - 2 \tau \zeta_x f^2(\alpha)\cos \alpha' \\
&&+ f^2(\alpha)- [1+  \tau\zeta_y f(\alpha)]^2=0 
\label{falpha.eqn}
\eea
where $f(\alpha)=(\frac{\eps+E_F}{\eps-E_F})\tau \sin\alpha \big/ [1+\tau(\zeta_x \cos\alpha+\zeta_y\sin\alpha)]$.
In the case $\zeta_x=0$ and $\eps\ll E_F$, this relation independent of the magnitude of $\zeta_y$
gives $\cos\alpha'=\cos\alpha$. On the other hand in the limit $\eps\ll E_F$ the conservation of $k_y$  
implies $\sin\alpha=-\sin\alpha$. Combining the two gives, $\alpha'=-\alpha$.
This means that in the $\eps\ll E_F$ regime, when the tilt is along the border, namely $\zeta_x=0$,
one recovers the {\em perfect retro} Andreev reflection. 
This is independent of the value of $\zeta_y$ and holds for any $\zeta_y$.
For generic tilt parameter ${\bs\zeta}=(\zeta_x,\zeta_y)$
the retro Andreev reflection will not satisfy the perfect retro reflection condition $\alpha'=-\alpha$. 

\begin{figure}[b]
\centering
\includegraphics[width=9cm]{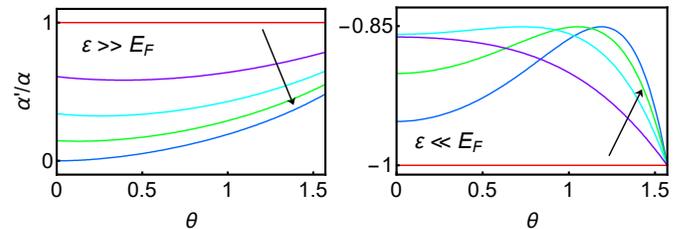}
\caption{(Color online) The ratio of the angle $\alpha'$ of the Andreev reflected hole to incident angle $\alpha$ of the electron.
The angle $\theta$ is the tilt angle. Various curves from red to blue correspond to $\zeta=n/4$ with $n=0$ to $4$. 
This figure is produced for $\alpha=\pi/6$. For all other angles the qualitative behavior is similar. In the
$\eps\gg E_F$ regime the Andreev reflection is specular-reflection (left), while in the opposite regime $\eps\ll E_F$
the Andreev reflection is retro-reflection (right). For $\zeta=1$ and $\eps\gg E_F$ (left) when the tilt is
perpendicular to the boundary, $\theta=0$, we have perpendicular Andreev reflection. 
}
\label{alpha_teta.fig}
\end{figure}

Let us investigate how does the tilt parameter $\bs\zeta$ affect the Andreev reflection.
Figure \ref{alpha_teta.fig} shows the ratio of the reflected hole angle ($\alpha'$) and incident electron angle ($\alpha$)
as a function of $\theta$, the angle of the tilt parameter $\bs\zeta$, namely $\zeta_y=\tan\theta \zeta_x$. 
The $\theta=0$ corresponds to a tilt in $x$ direction which is perpendicular to the N$|$S interface,
while $\theta=\pi/2$ corresponds to a tilt along the interface. 
Various colors from red to blue as indicated by the arrow direction increasingly correspond to $\zeta=0,0.25,0.5,0.75$ and $\zeta=1$.  
This color code holds through out the paper. 
This plot has been generated for two limits $\eps\gg E_F$ (left) and $\eps\ll E_F$ (right). 
As can be seen in the left panel corresponding to $\eps \gg E_F$, the ratio of the angles is positive.
This means that in this regime we only have the specular Andreev reflection. 
The red curve corresponds to $\zeta=0$ (i.e. without tilt). This part is in agreement with the earlier work 
of Beenakker~\cite{Beenakker_SAR}. As can be seen by moving from $\zeta=0$ (red curve) to $\zeta=1$ (blue curve)
the ratio of $\alpha'/\alpha$ is reduced below $1$. Therefore in the specular Andreev reflection dominated regime,
the effect of tilt is to reduce the angle $\alpha'$ of the specular reflected hole for any given incidence angle $\alpha$.
This means that the component of the current perpendicular to the interface is generically increased by tilting
the Dirac cone. For a given curve, the smallest specular reflection angle $\alpha'$ is obtained for $\theta=0$, i.e. for
the tilt perpendicular to the interface. It is interesting to note that for $\zeta=1$ for tilt along the $x$ direction (i.e. $\theta=0$),
the ratio of $\alpha'/\alpha$ is zero. This simply means that irrespective of the incidence angle of the electron,
the hole will be always reflected normal to the interface.

Now let us focus on the retro-Andreev regime of $\eps\ll E_F$ in the right panel of Fig.~\ref{alpha_teta.fig}. 
Again the red line corresponds to $\zeta=0$ where we have perfect retro Andreev reflection, namely $\alpha'/\alpha=-1$. 
For tilt along $y$ axis corresponding to $\theta=\pi/2$, as argued above for any tilt parameter $\zeta$ the above
perfect retro Andreev reflection condition in maintained. But for tilt angles to the left of $\pi/2$, by 
increasing the $\zeta$ from $0$ to $1$, the absolute value of the ratio decreases but still remains negative, 
meaning that we still have retro Andreev reflection which are not nevertheless perfect. 
In both panels the effect of tilt is to reduce the absolute value of the angle of the reflect hole -- i.e. absolute angle with respect to the interface normal. 

\begin{figure}[b]
\centering
\includegraphics[width=9cm]{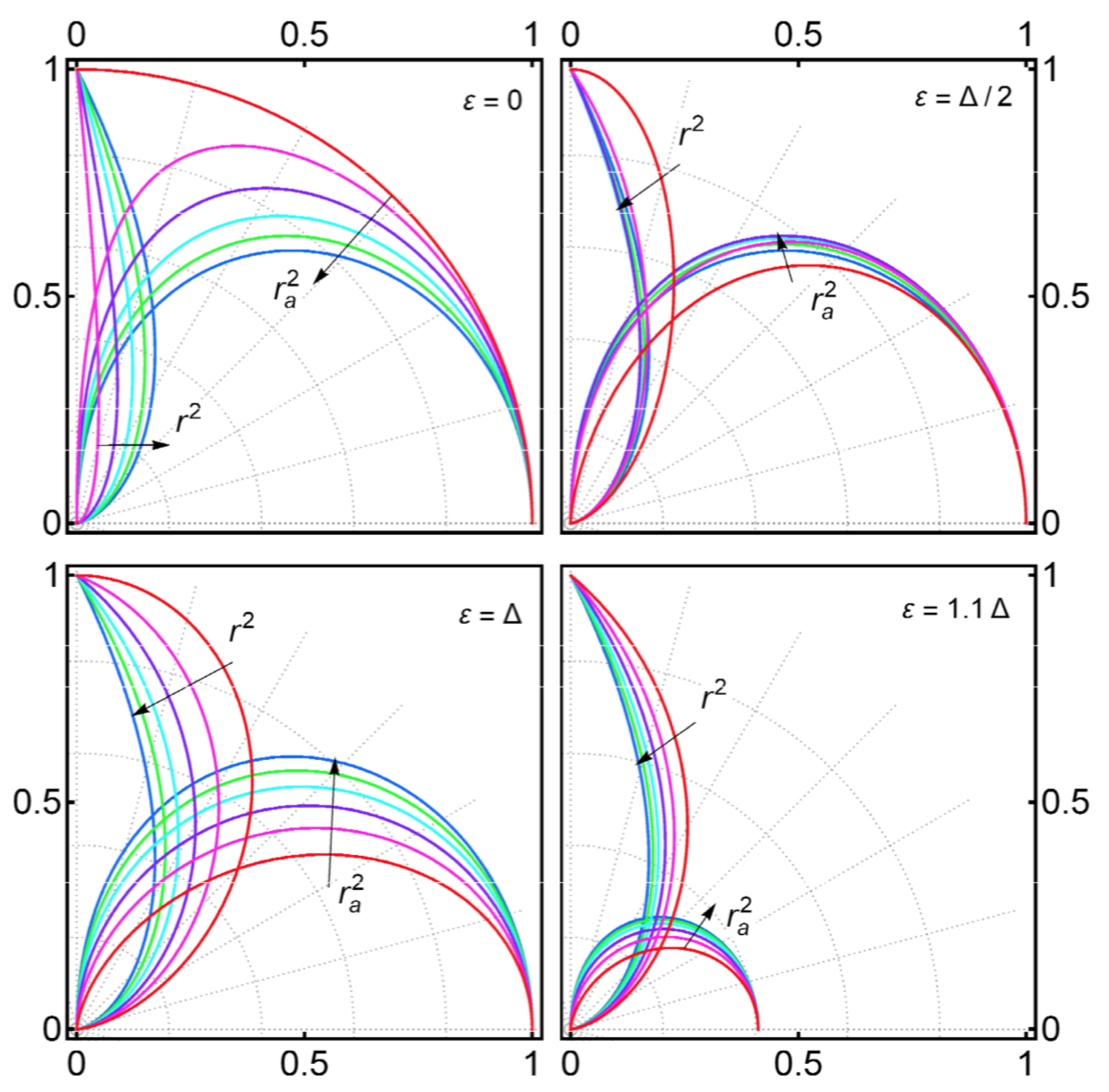}
\caption{(Color online) Polar plot of $r_a^2$ and $r^2$ in $\eps\gg E_F$ regime where SAR is dominant.
Color code is the same as in Fig.~\ref{alpha_teta.fig}.
Each panel corresponds to the value of $\eps$ indicated in the legend. 
The tilt is along the $x$ direction, namely $\theta=0$. The direction of arrows are from red with $\zeta=0$ to blue with $\zeta=1$.  
}
\label{r_ra_eg.fig}
\end{figure}

\section{Results}
\label{results.sec}
So far we have been focused on the angle of AR and its sign which provides information about the SAR (positive) and RAR (negative) processes. 
Now let us focus on the amplitude of Andreev reflection in the tilted Dirac cone system.
For this purpose, following Beenakker, we solve the scattering problem. 
Matching the wave functions in N and S regions and demanding continuity at $x = 0$, one can calculate the probability of the electron to hole conversion. 
We assume a large electrostatic potential $- U_0$ in S region (which can be adjusted by gate voltage or by doping). 
By this assumption, the eigenfunctions of the S region become independent of the tilt vector $\bs\zeta$ and simplify to,
\bea
\psi_S^\pm= e^{i (k_y y +k_x^\pm x)}
\big[ e^{\mp i \beta}, \pm e^{\mp i \beta}, e^{-i \phi}, \pm e^{-i \phi} \big]^{\rm T},
\eea
where superscript ${\rm T}$ indicates transpose, $\beta=\arccos (\eps / \Delta_0)$ if $\eps < \Delta_0$ and $-i~\text{arccosh} (\eps/\Delta_0)$ if $\eps > \Delta_0$. $\Delta_0$ and $\phi$ are the amplitude and phase of the superconductor and $k_x^{\pm}= \pm U_0/\hbar v - i (\Delta_0 / \hbar v) \sin \beta $. In this limit ($U_0 \gg E_F$ and $\eps$), the eigenstates of the S region are the same as tiltless Dirac equation~\cite{Beenakker_SAR}.
Continuity of the wave function across the NS boundary ($x=0$) gives the reflection amplitudes of an incident electron with angle $\alpha$ as follows:
\bea
\nn
&&r_a= \frac{  e^{-i \phi} \sqrt{\cos \alpha \cos \alpha'}}{\cos \beta \cos(\alpha' - \alpha)/2 + i \sin \beta \cos(\alpha'+\alpha)/2},\\
\nn
\\
&&r=\frac{- \cos \beta \sin(\alpha' + \alpha)/2 + i \sin \beta \sin(\alpha'-\alpha)/2}{\cos \beta \cos(\alpha' - \alpha)/2 + i \sin \beta \cos(\alpha'+\alpha)/2},
\eea 
where $r_a$ is the hole (Andreev) reflection amplitude and $r$ is the electron reflection amplitude. 

\begin{figure}[b]
\centering
\includegraphics[width=9cm]{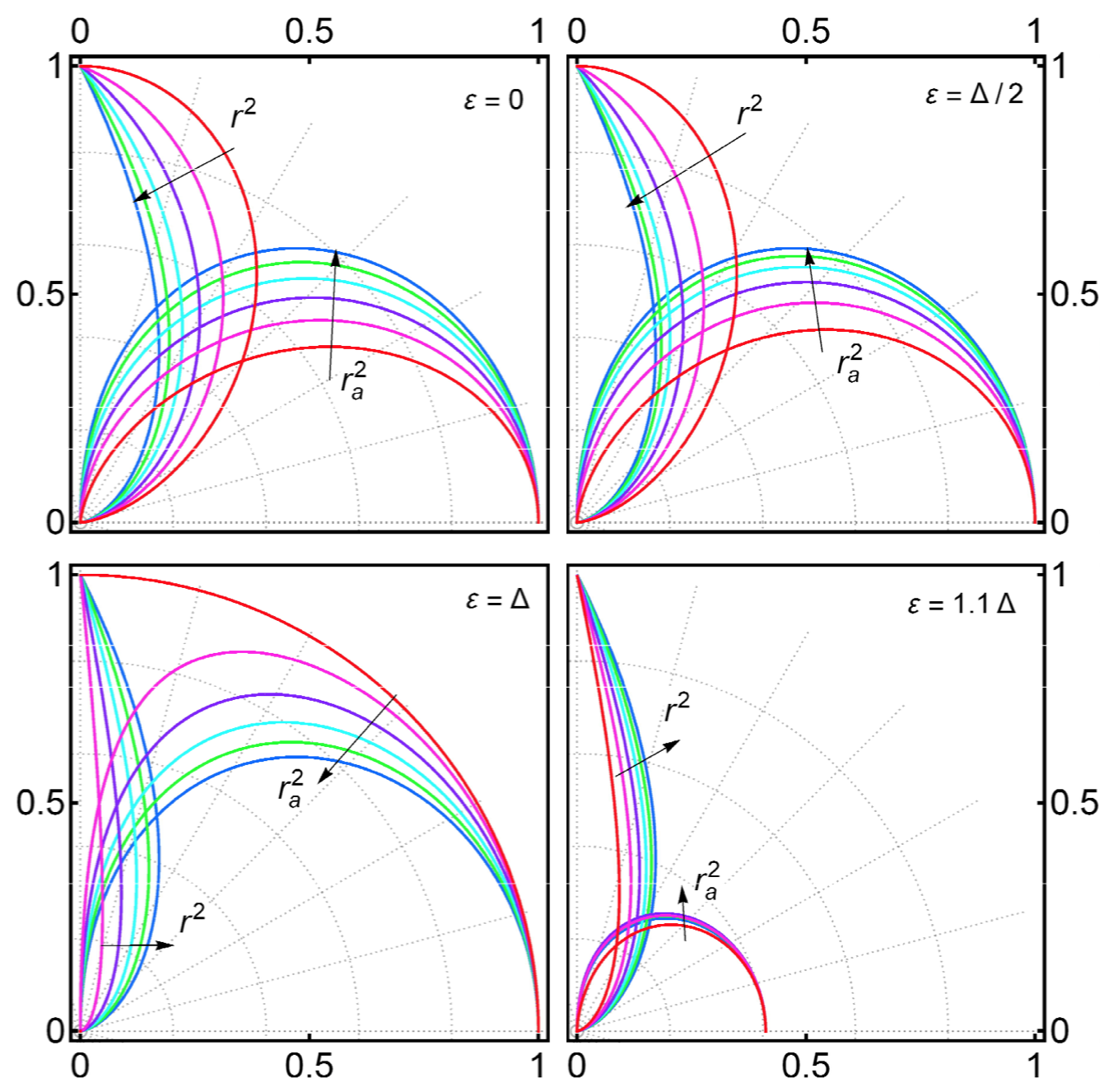}
\caption{(Color online) Same as Fig.~\ref{r_ra_eg.fig} in the RAR regime, $\eps\ll E_F$.}
\label{r_ra_el.fig}
\end{figure}
Fig.~\ref{r_ra_eg.fig} is a polar plot of the dependence of $r_a^2$ (Andreev reflection) and $r^2$ (standard reflection) for
the regime of $\eps\gg E_F$. The color code for the magnitude of tilt is same as in Fig.~\ref{alpha_teta.fig} and the tilt angle 
is $\theta=0$, i.e. the tilt is perpendicular to the interface. The energies are indicted in the legend of each panel. 
A common feature of all four panels is that the Andreev reflection becomes less by approaching $\alpha=\pi/2$
while the standard electron reflection is not possible for small angles around $\alpha=0$. That is why standard reflection curves are 
adjacent to $\alpha=\pi/2$ direction,
and Andreev reflection curves are accumulated around the $\alpha=0$ direction. 
Please note that the arc shaped curves of $r^2$ always extend up to $1$, while
$r_a^2$ can reach $1$ only for subgap energies. That is why in the bottom right panel
corresponding to $\eps=1.1\Delta$, the $r_a^2$ arcs do not extend up to $1$. 
This is a general feature, and holds in both SAR and RAR dominated regimes. 
Top left panel corresponds to $\eps=0$. It should be understood that in $\eps\gg E_F$ regime this means that
first the limit $E_F\to 0$ is taken and then we set $\eps=0$. In the top left panel, the red plot corresponding to $\zeta=0$
simply indicates that the Andreev reflection for all angles is $1$ and that the normal reflection is $r=0$ for all angles. 
Therefore one expects the $dI/dV=2$ for all angles which gives the average value of $2$ at $\eps=0$. This is in agreement
with Beenakker's result for upright Dirac cone. 
Except for $\eps=0$ where the red curve ($\zeta=0$) corresponds to larger $r_a^2$ than the other curves (i.e. non-zero $\zeta$s),
for other energies, the red curve is below the other curves. This means that for all angles, the Andreev reflection typically increases with 
increase of the tilt magnitude $\zeta$. 

In Fig.~\ref{r_ra_el.fig} we have plotted the same curves as Fig.~\ref{r_ra_eg.fig} in the $\eps\ll E_F$ regime.
This regime, as demonstrated in Fig.~\ref{alpha_teta.fig} corresponds to RAR. As for the
magnitudes of the reflections, again the general features are similar to Fig.~\ref{r_ra_eg.fig}. The major
difference for the subgap features is that the panels of the present figure from $\eps=\Delta$ down to $\eps=0$,
qualitatively behave similar to the $\eps=0$ up to $\eps=\Delta$ panel of Fig.~\ref{r_ra_eg.fig}. 

Having discussed the detailed dependence of $r^2$ and $r_a^2$ on the incidence angle $\alpha$, we 
are now ready to average over all the incidence directions and perform the angular integration to 
obtain the $dI/dV$ which at any given energy $\eps=eV$ is given by the BTK formula,
\be
   \frac{dI}{dV}=\frac{1}{g_0(eV)}\int_0^\infty dk_y \left(1-r^2+r_a^2 \right). 
   \label{BTK.eqn}
\ee

\begin{figure}[t]
\centering
\includegraphics[width=9cm]{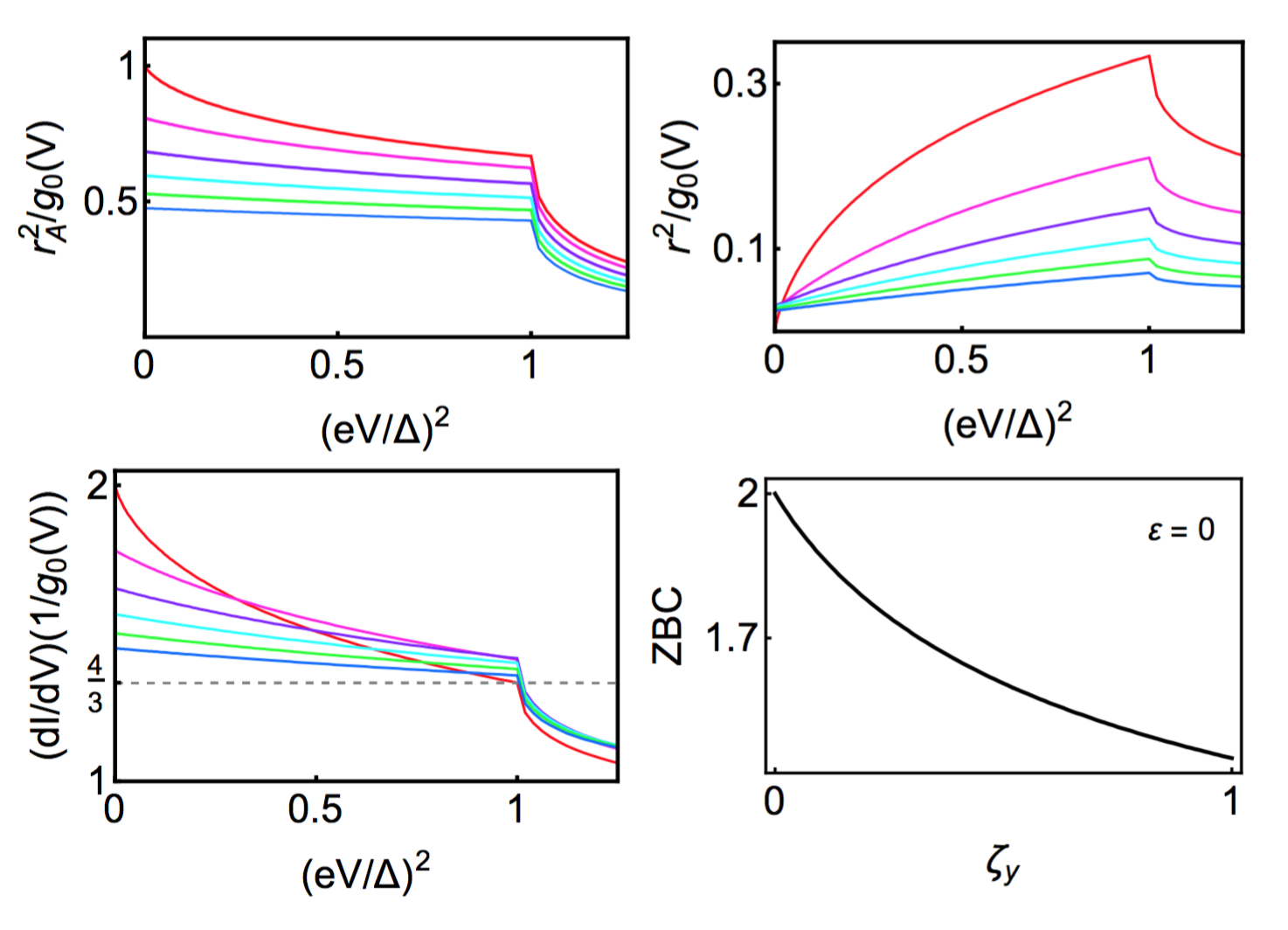}
\caption{(Color online) The color code is the same as Fig.~\ref{alpha_teta.fig}. 
Top left and right panels are the energy dependence of $r_a^2$ and $r^2$ of tilted Dirac fermions 
with tilt angle $\theta=\pi/2$ for various tilt parameters $\zeta$ in the SAR regime, $\eps\gg E_F$. 
Bottom-left panel is conductance of S$|$N junction and the bottom-right panel is the dependence of the zero-bias 
conductance to tilt parameter $\zeta_y$. 
}
\label{tetapi2_eg.fig}
\end{figure}
In this equation $g_0(eV)=(\frac{4e^2}{h})(E_F+\eps)W / \pi \hbar v$ normalizes the conductance. $W$ is the width of the Graphene sheet. 
In top row of Fig~\ref{tetapi2_eg.fig} we have plotted the dependence of $r_a^2$ (left) and $r^2$ (right) on energy for various tilts along $y$ direction. 
As we pointed out in the discussion below Eq.~\eqref{falpha.eqn}, when the tilts is along $y$ direction, the tilt does not play any role in 
the RARE regime, $\eps\ll E_F$.
That is why in this figure we have only plotted the $\eps\gg E_F$ (SAR) regime. The color code is as usual and from red to blue, $\zeta$ varies from $0$ to $1$. 
The generic line shapes for all colors are similar. The trend are also similar in both left and right panels: By increasing the tilt parameter $\zeta=\zeta_y$, both
$r_a^2$ and $r^2$ curves are pushed downward. From these information, one can extract the bottom row curves. In the bottom-left, using the BTK formula~\eqref{BTK.eqn}
we calculate the conductance profile for various values of $\zeta$. Since we need the difference between $r_a^2$ and $r^2$ in this formula, depending on which one 
is larger, the trend in conductance peak varies. Around $\eps=0$, by increasing the tilt, the conductance moves down, while at $\eps=\Delta$, by
increasing the tilt the conductance moves upward. In the bottom-right we have extracted the $\zeta_y$ dependence of the zero-bias conductance.
As can be seen, by increasing $\zeta$ it decreases.

\begin{figure}[t]
\centering
\includegraphics[width=9cm]{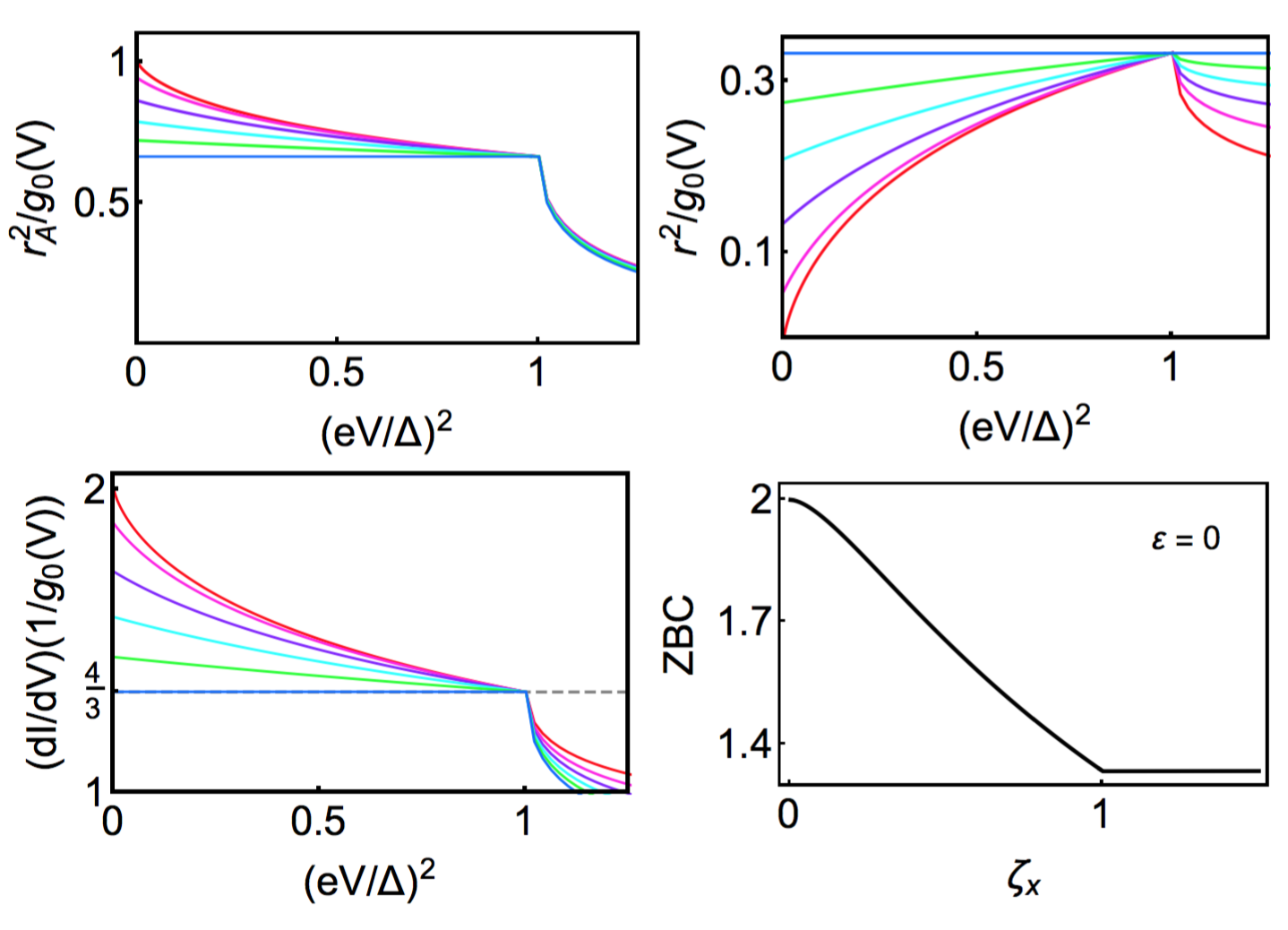}
\caption{(Color online) Same as Fig.~\ref{tetapi2_eg.fig} in the SAR regime $\eps\gg E_F$, but for tilt along the $x$ direction, namely $\theta=0$.   }
\label{teta0_eg.fig}
\end{figure}

Fig.~\ref{teta0_eg.fig} is the same as Fig.~\ref{tetapi2_eg.fig} for the tilt along $x$ axis, namely $\theta=0$. 
When the tilt is perpendicular to the interface, the $r_a^2$ curves all converge to a constant curve, $2/3$ (top left) at $\eps=\Delta$
while the $r^2$ curves converge to $1/3$ (top right) at this point. That is why the conductance in bottom-left 
converges to $1-1/3+2/3=4/3$ at the gap edge. In bottom-right panel we have extracted the zero-bias conductance as 
a function of $\zeta=\zeta_x$. This feature is similar to the one in Fig.~\ref{tetapi2_eg.fig}. 
A peculiar feature of the $\theta=0$ tilt is that for $\zeta=1$ (blue curve) both reflections will become
energy-independent in the subgap energies. That is why the conductance in bottom-left panel for $0\le\eps\le \Delta$
will be energy-independent. This energy independent subgap conductance can be regarded as a clear and robust 
signature of "horizon value", $\zeta=1$. 

\begin{figure}[t]
\centering
\includegraphics[width=9cm]{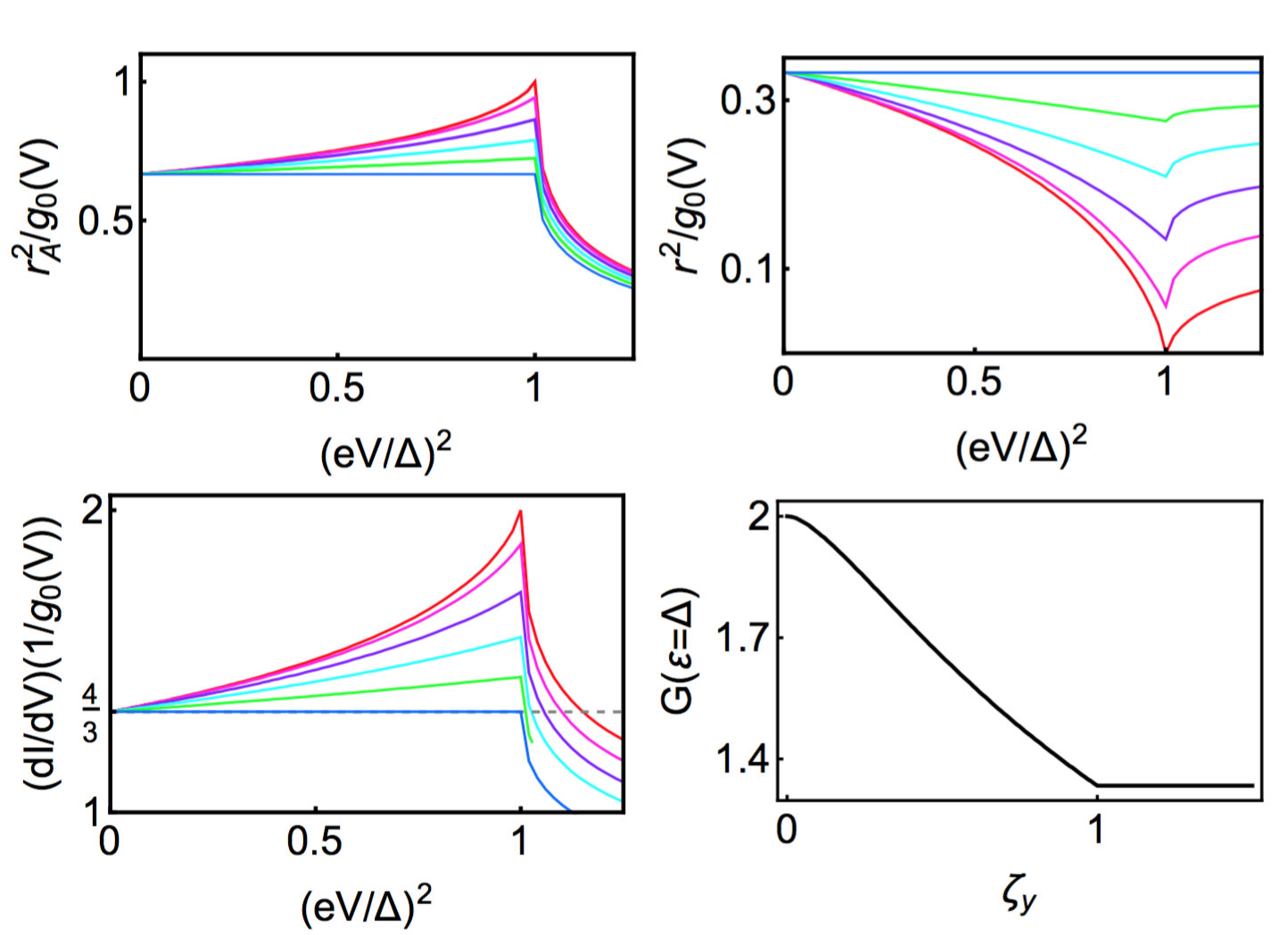}
\caption{(Color online) Same as Fig.~\ref{teta0_eg.fig} but for RAR regime, $\eps\ll E_F$. This corresponds to 
the regime of specular Andreev reflection.}
\label{teta0_el.fig}
\end{figure}

Fig.~\ref{teta0_el.fig} represents that same information as Fig.~\ref{teta0_eg.fig} for the RAR
regime, $\eps\ll E_F$. Again the lineshapes of $r_a^2$ and $r^2$ and the lines
move down by increasing $\zeta$ from $0$ (red) to $1$ (blue). The limiting $\zeta=1$ curves become flat
for both quantities in the subgap energies. The standard reflection lineshape becomes flat for all energies (top right),
however, the retro-Andreev reflection profile is energy independent only in the subgap region. 
For energies above the gap, Andreev reflections are suppressed and depend on energy. 
The conclusion from Fig.~\ref{teta0_eg.fig} (SAR) Fig. ~\ref{teta0_el.fig} (RAR) 
is that the conductance of N$|$S interface in the subgap region $0\le \eps\le \Delta$
is a constant when the tilt parameter corresponds to the horizon value, $\zeta=1$. 

\begin{figure}[b]
\centering
\includegraphics[width=9cm]{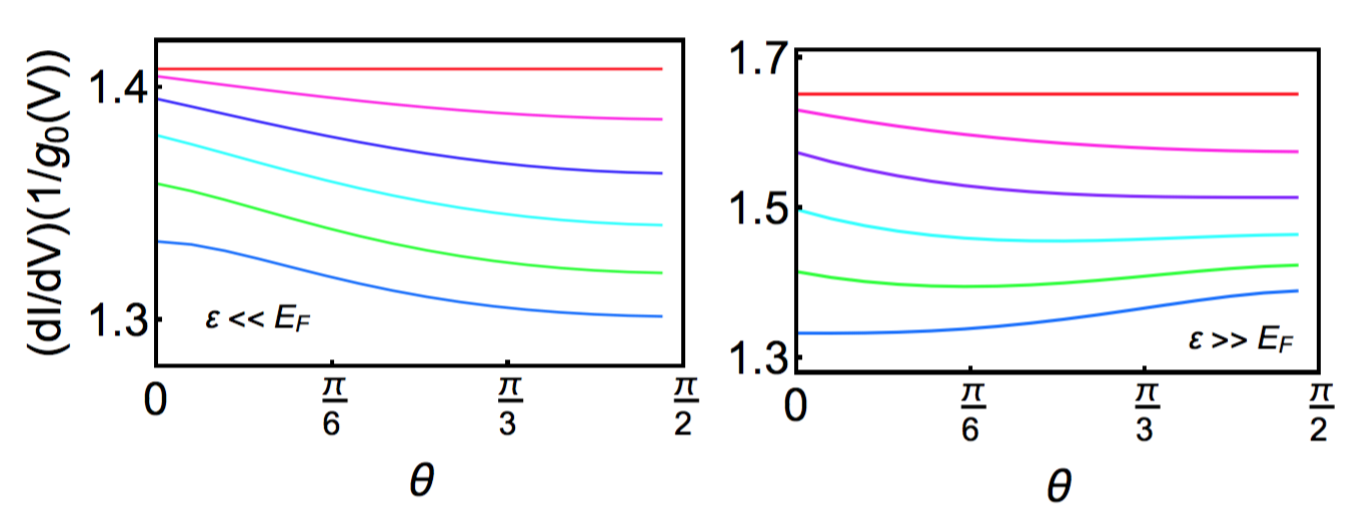}
\caption{(Color online) Dependence of conductance at subgap energy scale $\eps=\Delta/2$ on
the tilt angle $\theta$ for RAR regime, $\eps\ll E_F$ (left) and 
SAR regime, $\eps\gg E_F$ (right). 
The color code is the same as Fig.~\ref{alpha_teta.fig}. As can be seen
all curves corresponding to all values of $\zeta=0$ to $\zeta=1$ have very weak dependence 
on the tilt angle.}
\label{didv_teta.fig}
\end{figure}

So far we have investigated the dependence of Andreev and normal reflection on various
parameters for $\zeta=n/5$ where $n=0$ to $n=5$ correspond to red to blue curves.
The tilt direction characterized by $\theta$ has been fixed to be either in $x$ or
$y$ direction. How sensitive are the conductance data to the tilt angle?
To investigate this, in Fig.~\ref{didv_teta.fig} we have plotted the tilt angle
dependence of the conductance evaluated at a fixed subgap energy $\eps=\Delta/2$
for various tilt values $\zeta$. As can be seen the dependence on $\theta$ is very weak.
Therefore data represented in Figs.~\ref{teta0_eg.fig} and~\ref{teta0_el.fig} are 
typical behavior of reflection probabilities and the resulting conductances.

\section{Summary and outlook}
\label{discuss.sec}
In this paper we have studied the AR and standard reflection and conductance profile in tilted Dirac equation in 2+1 dimensions. 
In presence of tilt, again both RAR and SAR processes are possible. As the tilt $\zeta$
approaches $1$, both RAR and SAR will degenerate into PAR. This means that, there will 
be plenty of degrees of freedom which are forced to undergo nearly perpendicular AR as in Fig.~\ref{schematic.fig}. 
This can be clearly seen in Fig.~\ref{alpha_teta.fig}, where, by increasing $\zeta$ from $0$ to $1$ (corresponding to the
color code red to blue) the angle of the reflected hole comes closer to the perpendicular direction.
This is manifest for SAR regime of $\eps\gg E_F$. 
We found that for generic energies $\eps$ -- except for $\eps=0$ in the SAR regime and $\eps=\Delta$ in RAR regime --
and for all angles, when $\zeta$ increases from $0$ to $1$, the AR amplitude increases, while the standard reflection amplitude decreases. 

Despite the increase in angle resolved AR amplitudes, when it is integrated over all angles, due to
angular limitations, the overall $r_a^2$ decreases by increasing $\zeta$. This has been depicted in Figs.~\ref{tetapi2_eg.fig} and~\ref{teta0_eg.fig}. 
These effects leave a very clear signature of $\zeta$ in the conductance profile and the zero bias conductance peak in particular. 
The $\zeta\to 1$ limit is characterized by a flat subgap conductance profile in Figs.~\ref{tetapi2_eg.fig} and~\ref{teta0_eg.fig}
that correspond to two tilt angles $\theta=\pi/2$ (along $y$ direction) and $\theta=0$ (along $x$ direction). The
same flat conductance profile can be observed in RAR regime of Fig.~\ref{teta0_el.fig}. 
Finally in Fig.~\ref{didv_teta.fig} we demonstrated that the above effect does not depend much on 
the tilt angle in both SAR and RAR regimes. Therefore the conclusion is that, approaching the critically
tilted regime of $\zeta\to 1$ leaves a very clear signature in the transport measurements as a 
flat conductance profile in subgap energy scales. 

Let us speculate about a situation where $\bs\zeta$ depends on the coordinates:
In a geometric description of the tilted Dirac cone, the condition $\zeta=1$ corresponds to an event horizon~\cite{VolovikBH,Nissinen2017,Meng2018BH,tohid,Jalil2017,JafariEB}.
As we noted in Fig.~\ref{alpha_teta.fig}, the generic effect of increasing tilt parameter $\zeta$ (which geometrically means approaching to the solid-state 
horizon) is to bring the angle of Andreev reflected holes 
close to the interface normal. This in general enhances the Josephson current. Particularly for a tilt along $x$ direction 
as we saw in left panel of Fig.~\ref{alpha_teta.fig} when $\zeta=1$,
for any incident electron angle, the reflected hole is always reflected normal to the interface. 
Therefore the {\em perpendicular Andreev reflection} is associated with a black-hole horizon. 
This is neither specular, nor retro-Andreev reflection. The geometric interpretation of
perpendicular Andreev reflection is that in a curved solid-state spacetime~\cite{tohid,VolovikBH,Meng2018BH} admitting a horizon with $\zeta=1$, the Josephson
current will be maximized at the horizon. This is in agreement with earlier works on the enhancement of 
pairing correlations~\cite{Volovik_2018,Alidoust2019evolution} at $\zeta=1$. When the $\bs\zeta$ is constant 
all over the spacetime, such enhancement can be attributed understood as accumulation of the density of states
around the Fermi level. But the present geometric construction relating the enhancement of superconducting
correlations to the presence of horizon is much more general and applies to spacetime dependent $\bs\zeta$ case too. 

Our finding of geometry dependent conductance profile suggests the S$|$N conductance measurements
as a useful spectroscopic tools to investigate the geometry of the solids with alternative spacetimes. 
Extension of the present work to local scanning tunneling microscopy with superconducting tips 
to probe local geometry of the alternative spacetims in the solids is desirable. 

\section{Acknowledgements}
S. A. J. was supported by grant No. G960214 from the research deputy of Sharif University of Technology
and Iran Science Elites Federation (ISEF). Z. F. was supported by a post doctoral fellowship from ISEF.
S. A. J. is grateful for the hospitality of Perimeter Institute where part of this work was carried out.  
Research at Perimeter Institute is supported in part by the Government of Canada through the Department 
of Innovation, Science and Economic Development Canada and by the Province of Ontario through the Ministry of 
Economic Development, Job Creation and Trade.
Z. F. is grateful to the Abdus Salam center for Theoretical Physics for a long term visit during which this research was completed.

\bibliography{mybib}

\end{document}